\documentclass[12pt]{article}
\usepackage[cp1251]{inputenc}
\usepackage{epsf}

\title{Valence quark distributions in nucleon at low $Q^2$ in QCD.}
\author{B.L.Ioffe and A.G.Oganesian\\
Institute of Theoretical and
Experimental Physics,\\
B.Cheremushkinskaya 25, 117218 Moscow,Russia}
\date{}
\begin{document}
\maketitle

\newcommand{\be}{\begin{equation}}
\newcommand{\ee}{\end{equation}}

\def\la{\mathrel{\mathpalette\fun <}}
\def\ga{\mathrel{\mathpalette\fun >}}
\def\fun#1#2{\lower3.6pt\vbox{\baselineskip0pt\lineskip.9pt
\ialign{$\mathsurround=0pt#1\hfil##\hfil$\crcr#2\crcr\sim\crcr}}}

\begin{abstract}
Valence $u$- and $d$-quarks distributions in proton are calculated in QCD at
low $Q^2$ and intermediate $x$, basing on the operator product expansion
(OPE). The imaginary part of the virtual photon scattering amplitude on
quark current with proton quantum numbers is considered. The initial and
final virtualities $p^2_1$ and $p^2_2$ of the currents are assumed to be
large, negative and different, $p^2_1 \not= p^2_2$. The OPE in $p^2_1$,
$p^2_2$ up to dimension 6 operators was performed. Double dispersion
representations in $p^2_1, p^2_2$ ~ of the amplitudes in terms of physical
states contributions are used. Putting them to be equal to those calculated
in QCD, the sum rules for quark distributions are found. The double Borel
transformations are applied to the sum rules. Leading order perturbative
corrections are accounted. Valence quark distributions are found: $u(x)_v$
at $0.15<x<65$,~ $d(x)_v$ at $0.25<x<0.55$ with an accuracy $\sim 30\%$ in
the middles and $\sim 50\%$ at the ends of these intervals. The quark
distributions obtained are in agreement with those found from the analysis
of hard processes data.
\end{abstract}

\vspace{5mm}


\vspace{3mm}


\vspace{1cm}

\begin{center}
{\bf 1.Introduction}
\end{center}

\vspace{5mm}
Quark and gluon distributions in hadrons are not fully understood in QCD.
QCD predicts the evolution of these distributions with $Q^2$ in accord with
the Dokshitzer-Gribov-Lipatov-Altarelli-Parisi (DGLAP) [1]-[3] equations,
but not the initial values from which this evolution starts. The standard
way of determination of quark and gluon distributions in nucleon is the
following [4-8] (for the recent review see [9]). At some $Q^2 = Q^2_0$
(usually, at low or intermediate $Q^2 \sim 2-5 GeV^2$) the form of quark
(valence and sea) and gluon distributions is assumed and characterized by
the number of free parameters. Then, by using DGLAP equations, quark and
gluon distributions are calculated at all $Q^2$ and $x$ and compared with
the whole set of the data on deep inelastic lepton-nucleon scattering
(sometimes also with prompt photon production, jets at high $p_{\bot}$ etc).
The best fit for the parameters is found and, therefore, quark and gluon
distributions are determined at all $Q^2$, including their initial values
$q(Q^2_0, x)$, ~$g(Q^2_0, x)$. Evidently, such an approach is not completely
satisfactory from theoretical point of view - it would be desirable to
determine the initial distribution directly from QCD. Also, if the form of
initial distribution is not divined properly, then the fitting procedure
will give the relative minimum, but not the absolute ones and the results,
especially at low $Q^2_0$ could be wrong. (This danger does not, probably,
exists for nucleon, but such a situation happens, e.g., for transversally
polarized $\rho$-meson, where unusual form of initial valence quark
distribution was found [10]). Finally, extrapolation from high $Q^2$, to
which the main part of the data belongs, to low $Q^2$ is not a quite stable
procedure and may introduce some errors. For all these reasons it is
desirable to find quark and gluon distribution in hadrons at low $Q^2 \sim
2-5 GeV^2$ basing directly on QCD.

In this paper we calculate $u$ and $d$ valence quark distributions in
proton. The idea of the method was suggested in [11] and developed in
[12-14]. Recently, the method  had been improved and valence quark
distributions in pion [15] and transversally and longitudinally polarized
$\rho$-meson [10] had been calculated, what was impossible in the initial
version of the method. The idea of the approach (in the improved version) is
to consider the imaginary part (in $s$-channel) of a four-point correlator
$\Pi(p_1, p_2, q, q^{\prime})$ corresponding to the non-forward scattering
of two quark currents, one of which has the quantum numbers of hadron of
interest (in our case -- of proton) and the other is electromagnetic (or
weak). It is supposed that virtualities of the photon  $q^2, q^{\prime 2}$
and hadron currents $p^2_1, p^2_2$ are large and negative $\vert q^2 \vert =
\vert q^{\prime 2} \vert \gg \vert p^2_1 \vert,~ \vert p^2_2 \vert \gg
R^{-2}_c$, where $R_c$ is the confinement radius. It was shown in [12] that
in this case the imaginary part in $s$-channel $[s = (p_1 + q)^2]$ of
$\Pi(p_1, p_2; q_1, q^{\prime})$ is dominated by a small distance
contribution at intermediate $x$. (The standard notation is used: $x$ is the
Bjorken scaling variable, $x = -q^2/2 \nu$~, $\nu = p_1 q)$. The proof of
this statement is given in ref.[12]. So, in the mentioned above domain of
$q^2, q^{\prime  2}$,~ $p^2_1, p^2_2$ and intermediate $x$~ $Im \Pi(p,p_2;
q, q^{\prime})$ can be calculated using the perturbation theory and the
operator product expansion in both sets of variables $q^2 = q^{\prime 2}$
and $p^2_1, p^2_2$. The approach is inapplicable at small $x$ and $x$ close
to 1. This can be easily understood for physical reasons.  In deep inelastic
scattering at large $\vert q^2 \vert$ the main interaction region in
space-time is the light-cone domain and longitudinal distances along the
light-cone are proportional to $1/x$ and become large at small $x$ [16, 17].
For OPE validity it is necessary for these longitudinal distances along
light-cone to be also small, that is not the case at small $x$. At $1 - x
\ll 1$ another condition of  applicability of the method is violated. The
total energy square $s = Q^2(1/x - 1) + p^2_1$~ $Q^2 = -q^2$ is not large at
$1 - x \ll 1$. Numerically, the typical values to be used below are  $Q^2
\sim 5 GeV^2$, ~ $p^2_1 \sim - 1 GeV^2$. Then, even at $x \approx 0.7$,~~ $s
\approx 1 GeV^2$, i.e., at such $x$ we are in the resonance, but not in the
scaling region. So, one may expect beforehand, that our method could work
only up to $x \approx 0.7$. The inapplicability of the method at small and
large $x$ manifests itself in the blow-up of higher order terms of OPE. More
precise limits on the applicability domain in $x$ will be found from the
magnitude of these terms.

The further procedure is common for QCD sum rules. On one hand the four-point
correlator $\Pi(p_1, p_2; q, q^{\prime})$ is calculated by perturbation
theory and OPE.On the other hand, the double dispersion representation in
$p^2_1, p^2_2$ in terms of physical states contributions is written for the
same correlator and the contribution of the lowest state is extracted using
the Borel transformaion. By equalling these two expression the desired quark
distribution is found.

In Sec.2 the general outline of the method is presented. It is explained
here, why it is necessary to consider at the beginning the nonequal
$p^2_1$ and $p^2_2$ and only at the end of the calculation to go to the
forward scattering amplitude. Sec.3 presents the results of the bare loop
calculation ($d = 0$ term in OPE) and the LO perturbative corrections to it.
Sec.4 is devoted to the calculation of power corrections to the QCD side of
the sum rules: the contribution of gluonic condensate (d=4), $\alpha_s
\langle \bar{\psi} \psi \rangle^2$ term (d=6).
The valence $u$ and $d$-quark
distributions are numerically calculated in Sec.5, the errors are estimated
and the results are compared with those obtained from the deep
inelastic scattering data using the evolution equations. Sec.6 contains our
conslusion.

\vspace{1cm}

\begin{center}
{\bf 2. The outline of the method}
\end{center}

\vspace{3mm}

Consider the 4-current correlator which corresponds to the virtual photon
scattering on the quark current with quantum number of proton:

$$
T^{\mu \nu} (p_1, p_2, q, q^{\prime}) = - i \int d^4 x d^4 y d^4 z \cdot
e^{i(p_1 x+ q y - p_2 z)} $$
\be
\cdot \langle 0 \vert T \{ \eta (x),~
j^{u, d}_{\mu} (y), j^{u, d}_{\nu} (0), ~ \bar{\eta} (z) \} \vert 0 \rangle,
\ee

where $\eta(x)$ is the three-quark current (Ioffe current [18]).
Choose the currents in the form $j^u_{\mu} = \bar{u}
\gamma_{\mu}u$,~ $j^d_{\mu} = \bar{d} \gamma_{\mu} d$, i.e. as an
electromagnetic current which interacts only with  $u(d)$ quark
(with unit charges). Such a choice allows us to get sum rules
separately for distribuion functions of $u$ and $d$ quarks. The
general method of calculation of distribution functions from
consideration of the 4-point correlator in the QCD sum rules for
light quarks was used in [12]. Unfortunately, this approach being
in the form used in [12] did not give completely satisfactory sum
rules for quark distribution functions (especially for $d$-quark
in nucleon) and did not allow one to calculate quark distributions
in mesons. Recently, in ref. [10,15] was suggested a generlized
method of calculation of quark distribution functions, which made
it possible to get much more reliable sum rules, what, in
particular, made it possible to find valence quark distributions
in $\pi$ and $\rho$- mesons.  The main difference of this method
is that the hadronic current momenta are put to be unequal, $p^2_1
\not= p^2_2$ and then independent borelization over $p^2_1$ and
$p^2_2$ is performed and only at the very end the Borel the
parameters $M^2_1$ and $M^2_2$ are put to be equal. (Unlike the
approach of ref.[12] where $p_1$ was put to be equal to $p_2$ and
single borelization is performed from the very beginning). The
described procedure allows one to kill nondiagonal transitions of
the type

\be
\langle 0 \vert j^h \vert h^* \rangle \langle h^* \vert j^{el}_{\mu} (y)
j^{el}_{\nu} (0) \vert h \rangle \langle h \vert j^h \vert 0 \rangle
\ee
and thus makes it possible to
separate the diagonal transition of interest

\be
\langle 0 \vert j^h \vert h \rangle \langle h \vert j^{el}_{\mu} (y)
j^{el}_{\nu} (0) \vert h \rangle \langle h \vert j^h \vert 0 \rangle,
\ee
without using additional tricks like differentiation over Borel parameter
$M^2$, what strongly worsens the accuracy of the sum rules.

Let us briefly remind the main points of derivation of the sum
rules in the method under investigaion (for details see [10,
15]. The most general form of the double dispersion relation
(in $p^2_1$ and $p^2_2$) for imaginary part of the correlator
(1) (we omit for some time all indeces assuming the necessary
invariant amplitude to be chosen) has the form
 $$
 Im T (p^2_1, p^2_2, q^2, s) = a(q^2, s) + \int\limits^{\infty}_{0}
 \frac{\psi(q^2, s, u)}{u - p^2_1} du + \int\limits \frac{\psi(q^2, s, u)}{u
 - p^2_2} du
 $$
 \be
 + \int\limits^{\infty}_{0}~ du_1 ~ \int\limits^{\infty}_{0}~ du_2~
 \frac{\rho(q^2, s, u_1, u_2)}{(u_1 - p^2_1)(u_2 - p^2_2)}
 \ee
(without loss of generality
one may put $q^2 = q^{\prime 2}$,  ~ $t
 = (p_1 - p_2)^2 = 0$).
 The double Borel transformation in $p^2_1$ and $p^2_2$ eliminates three
 first terms and we have

 \be
B_{M^2_1} B_{M^2_2} Im T (p^2_1, p^2_2, q^2, s) = \int\limits^{\infty}_{0}
du_1 ~ \int\limits^{\infty}_{0}~ du_2 \rho (q^2, s, u_1, u_2) exp \Biggl [ -
\frac{u_1}{M^2_1} - \frac{u_2}{M^2_2} \Biggr ]
\ee
where $M^2_1$ and $M^2_2$ are the squared Borel mass. The integration region
with respect to $u_1, u_2$ may be divided into four areas

I. $u_1 < s_0$,~ $u_2 < s_0$;

II. $u_1 < s_0$, ~ $u_2 > s_0$;

III. $u_1 > s_0$, ~ $u_2 < s_0$;

IV. $u_1$, $u_2 > s_0$.

Here $s_0$ is the continuum threshold in the standard QCD sum rule model of
the hadronic spectrum with one lowest resonance plus continuum. Area I
obviously corresponds to the resonance contribution and spectral density in
this area can be written as

\be
\rho(u_1, u_2, x, Q^2) = g^2_h \cdot 2 \pi F_2(x, Q^2) \delta(u_1 - m^2_h)
\delta(u_2 - m^2_h),
\ee
where $g_h$ are coupling constants of the corresponding hadronic current. As
for the areas II-IV, they are exponentially suppressed, and, using the
standard hypothesis of quark-hadron duality, we may estimate them as a
bare loop contribution in the same integration region.

Thus, the general form of the sum rule is

$$
Im T^0_{QCD} + \mbox{Power correction} = 2 \pi F_2 (x, Q^2) g^2_h
e^{-m^2_h(\frac{1}{M^2_1}+\frac{1}{M^2_2})}
$$
\be
Im T^0_{QCD} = \int\limits^{s_0}_{0}~ \int\limits^{s_0}_{0}~ \rho^0(u_1,
u_2, x) du_1 du_2 e^{-(\frac{u_1}{M^2_1}+\frac{u_2}{M^2_2})}
\ee

where one should go to the limit $M^2_1 = M^2_2$.

We choose $M^2_1 = M^2_2 = 2 M^2$ following [19] where it was shown that
the value of the Borel mass square in the double sum rules is approximately
twice as large than in single ones.

Choose now invariant amplitude for the case of the proton current of
interest in the correlator (1) (cf.[12]). Its imaginary (in $s$) part can be
written as

\newpage

$$
Im T^{(p)}_{\mu \nu} = \lambda^2_n \frac{1}{p^2_1 - m^2} ~~ \sum\limits_{r,
r^{\prime}}~ v^r(p_1) \times
$$
\be
\times Im \{ -i  \int d^4 x e^{-i q y} \cdot
\langle p_1, r \vert T \{j_{\mu}(y), j_{\nu} (0) \} \vert p_2,
r^{\prime} \rangle  \}  \cdot \bar{v}^{r^{\prime}}(p_2)
\frac{1}{p^2_2 - m^2}
\ee
where $v^r(p)$ is proton spinor with polarization $r$ and momentum $p$,
$\lambda_N$ is the coupling constant of proton with the current $\langle 0
\vert \eta \vert p, r \rangle = \lambda_N v^r(p)$.

In order to choose the most suitable invariant amplitude, rewrite eq.(8) in
the form

$$
Im T_{\mu \nu} = \lambda^2_N \frac{1}{(p^2_1- m^2)(p^2_2 - m^2)}
\sum\limits_{r, r^{\prime}}~ v^r(p_1) \cdot
$$
\be
\cdot \Biggl (\bar{v}^r(p_1) Im~\tilde{T}^{(p)}_{\mu \nu}~
v^{r^{\prime}} (p_2) \Biggr ) \bar{v}^{r{\prime}} (p_2)
\ee
where
$Im~\tilde{T}^{(p)}_{\mu \nu}$ is the amplitude before averaging in proton
spin, $m$ is the proton mass.

The general form of this amplitude is

$$
Im~\tilde{T}^{(p)}_{\mu \nu}(p_1, p_2) = C_1 P_{\mu} P_{\nu} + C_2(P_{\mu}
\gamma_{\nu} + P_{\mu} \gamma_{\nu}) + C_3 \hat{P} P_{\mu} P_{\nu} + ...
$$
\be
+ (\mbox{terms with}~ r)
\ee

where $P = \frac{p_1 + p_2}{2}$; ~~ $r = p_1 - p_2$.

Let us now take into account that we are interested in such a combination
of invariant amplitudes which in the spin-averaged matrix element

\be
\sum\limits_{r} \bar{v}^r (p_1) Im~\tilde{T}^{(p)}_{\mu \nu} v^r (p_2)
\ee
appears at the kinematical structure $P^{\mu}P^{\nu}$ (since this structure
in the limit $p_1 \to p_2 \equiv p$ transforms into $p^{\mu} p^{\nu}$, the
coefficient at which is just $F_2(x)$).

Using the equation of motion

\be
\hat{p}_{1,2} v^r(p_{1,2}) = \frac{1}{2} (\hat{p}_{1,2} + m) v^r(p_{1,2})
\ee
and
\be
\sum\limits_{r} v^r_{\alpha} (p_{1,2}) \bar{v}^r_{\beta} (p_{1,2}) =
(\hat{p}_{1,2} + m)_{\alpha \beta}
\ee
it can be seen that the combination of invariant amplitudes in eq.(10),
which appears in eq.(9) at the
kinematical structure $\hat{P} P^{\mu} P^{\nu}$ coincides (up to numerical
factor) with the combination of invariant functions at the structure
$P_{\mu}P_{\nu}$ in (11).

Thus, we come to a conclusion that the sum rules should be written for
invariant amplitude at the kinematical structure $\hat{P} P^{\mu} P^{\nu}$
(in what follows we will denote it $Im T/\nu$).

So, the sum rules for nucleon have the form

\be
\frac{2 \pi}{4 M^4}~ \frac{\bar{\lambda}^2_N}{32 \pi^4} ~ x q^{u,d} (x)
e^{-m^2/M^2} = Im T^0_{u,d} + \mbox{Power~ corrections}
\ee
where $\bar{\lambda}^2_N = 32 \pi^4 \lambda^2_N$; ~ $q^{u,d}(x)$ are
distribution functions of $u(d)$ quark in nucleon, $Im \tilde{T}^0$ is
perturbative contribution, i.e. of a bare loop with perturbative
corrections. (The continuum contribution, i.e., of II, III, IV regions
should be subtracted from $Im \tilde{T}$. Note, that really, the
contribution of regions II and III to the bare loop is zero, since $\rho^0
\sim \delta(u_1 - u_2)$).

\vspace{1cm}

\begin{center}
{\bf 3. Bare loop contribution and leading order perturbative corrections.}
\end{center}

\vspace{3mm}

Bare loop contribution to the sum rules is represented in Fig.1. In the
calculation we use the technique described in ref.15, Appendix. The
following formulae are exploited:

$$
\int~ \frac{d^4 k}{(p_1 - k)^2 (p_2 - k)^2} \delta [(p_1 + q - k)^2]
\theta(k^2) = \frac{\pi}{4 \nu x} (1 - x) \int~ du \frac{u}{(p^2_1 -
u)(p^2_2 - u)}
$$
$$
\int~ \frac{d^4 k \cdot k^2}{(p_1 - k)^2 (p_2 - k)^2} \delta [(p_1 + q -
k)^2] \theta(k^2) = \frac{\pi}{8 \nu x} (1 - x)^2~ \int~ du
\frac{u^2}{(p^2_1 - u)(p^2_2 - u)}
$$
\be
\int~ \frac{d^4 k(p_1-k)p_2}{(p_1 - k)^2 (p_2 - k)^2} \delta [(p_1 + q
- k)^2] \theta(k^2) = \frac{\pi}{16 \nu x} (1 - x^2)~ \int~ du
\frac{u^2}{(p^2_1 - u)(p^2_2 - u)}
\ee
(The terms, vanishing at the double Borel transformation are omitted). The
results after the double Borel transformation are the same as in the case for
equal $p_1 = p_2$ [12]:

\be
Im T^0_{u(d)} = \varphi^{u(d)}_0 (x) \frac{M^2}{32 \pi^3}~
E_2 (s_0/M^2)
\ee
where

\be
\varphi^u_0 (x) = x(1-x)^2 (1+8 x), ~~ \varphi^d_0(x) = x(1-x)^2 (1-2x),
\ee

$s_0$~~ \mbox{is ~ the~ continuum~ threshold}

\be
E_2(z) = 1 - (1+z+z^2/2)e^{-z}
\ee
The substitution of eq.(16) into the sum rules (14) results in

\be
x q(x)_0^{u(d)} = \frac{2 M^6
e^{m^2/M^2}}{\bar{\lambda}^2_N}~\varphi_0^{u(d)}(x) \cdot E_2
(\frac{s_0}{M^2})
\ee
In the bare loop approximation the moments of the quark structure function
are equal to
$$
\int\limits^{1}_{0}q^d_0(x) dx = \frac{M^6 e^{m^2/M^2}}{\bar{\lambda}^2_N}
E_2
$$
\be
\int\limits^{1}_{0}~ q^u_0(x) dx = 2 \frac{M^6
e^{m^2/M^2}}{\bar{\lambda}^2_N} E_2
\ee
Making use of relation
$\bar{\lambda}^2_N e^{-m^2/M^2} = M^6 E_2$ which follows from the sum rule
for the nucleon mass (see [18]) in the same approximation, we get

$$
\int\limits^{1}_0~q^d_0(x) dx = 1
$$
\be
\int\limits^{1}_0~q^u_0(x) dx = 2
\ee
In the bare loop approximation there also appears the sum rule for the
second moment:
\be
\int\limits^{1}_{0}~ x(q^u_0(x) + q^d_0(x)) dx = 1
\ee

Analogously to [12] one can show that relations (21),(22) hold also when
taking into account power corrections proportional to the quark condensate
square in the sum rules for the 4-point correlator (Fig.2)  and in the sum
rules for the nucleon mass. Relations (21) reflect the fact that proton has
two $u$-quarks and one $d$-quark. Relation (22) expresses the momentum
conservation law -- in the bare loop approximation all momentum is carried
by valence quarks. Therefore, the sum rules (21),(22) demonstrate that the
zero order approximation is reasonable. In the real physical theory the
regions $x \ll 1$ and $1-x \ll1$ are off the frames of our consideration.
However, in the noninteracting quark model which corresponds to the bare
loop approximation, the whole region $0 \leq x \leq 1$ should be considered
and relations (21),(22) should take place.

Let us calculate the perturbative corrections to bare loop and restrict
ourselves by the leading order (LO) corrections proportional to $ln
Q^2_0/\mu^2$, where $Q^2_0$ is the point, where the quark distributions
$q(x, Q^2_0)$ is calculated and $\mu^2$ is the normalization point. In our
case it is reasonable to choose $\mu^2$ to be equal to the Borel parameter
$\mu^2 = M^2$. The results take the form:

$$
d^{LO}(x) = d_0(x) \left \{1+\frac{4}{3} ln (Q^2_0/M^2) \cdot
\frac{\alpha_s(Q^2_0)}{2 \pi} \cdot \right.
$$
\be
\left. \Biggl
[1/2+x+ln((1-x)^2/x) + \frac{-5-17x+16x^2+12x^3} {6(1-x)(1+2x)} -
\frac{(3-2x)x^2 ln (1/x)}{(1-x)^2(1+2x)} \Biggr ] \right \}
\ee

\vspace{3mm}

$$
u^{LO}(x) = u_0(x) \cdot  \left \{
1+\frac{4}{3}~\frac{\alpha_s(Q^2_0)}{2 \pi}ln (Q^2_0/M^2) \Biggl [ 1/2+x+ln(1-x)^2/x +  \right.
$$
\be
\left. \frac{7-59x+46x^2+48x^3}{6(1-x)(1+8x)} - \frac{(15-8x)x^2 ln
(1/x)}{(1-x)^2(1+8x)} \Biggr ] \right \}
\ee
where $u_0(x)$ and $d_0(x)$  are bare loop contributions, given by (19).

\newpage

\begin{center}
{\bf 4.Power corrections}
\end{center}

\vspace{3mm}
In this Section the power corrections to the sum rules are calculated.

The lowest dimension (d=4) power corrections are the corrections due to
gluon condensate $\langle 0 \vert(\alpha_s/\pi) G^n_{\mu \nu}~G^n_{\mu \nu}
\vert 0 \rangle$. Some examples of the corresponding diagrams are given in
Fig.3. The calculations are performed in the fixed point (Fock-Schwinger)
gauge. The corresponding expression for quark propagator is given in ref.12,
eq.(24). (The error in the coefficient in front of the last term was
corrected: it should be 1/288 instead of 1/96). In order to be sure in the
final results the fixed point was chosen in two ways: at the upper and lower
left-hand vertices of Fig.1 diagrams. The results for the sum of diagrams
coincide in these two cases, as it should be. However, the contributions of
separate diagrams are different. Particularly, if the fixed point is chosen
at the upper left-hand vertex, then all diagrams where the soft gluon is
emitted from the upper horizontal line (i.e., by active quark) in Fig.1, are
zero in the lowest twist approximation. However, if the fixed point is at
the lower left-hand vertex, the diagrams with soft gluon, emitted by active
quark are nonzero.  \footnote{May be, this observation is related
to the Brodsky et al. [20] statement, that structure functions are
influent by active quark interactions with spectators.} Therefore, it is
generally untrue the folklore statement, that in the lowest twist
approximation one may neglect in the forward scattering amplitude the
active quark interaction between absorption and emission of virtual photon
-- this statement is gauge dependent.

All other technique is the same as in ref.15. The contributions of gluon
condensate to $u$ and $d$-quarks distribution were found to be (the ratios
to bare loop contributions are presented):

\be
\frac{u(x)_{\langle G^2 \rangle}}{u_0(x)} = \frac{\langle (\alpha_s/\pi) G^2
\rangle}{M^4} \cdot \frac{\pi^2}{12}~\frac{(11+4x-31x^2)}{x(1-x)^2(1+8x)}
\cdot (1-e^{-s_0/M^2})/E_2(\frac{s_0}{M^2})
\ee

\vspace{3mm}

\be
\frac{d(x)_{\langle G^2 \rangle}}{d_0(x)} = - \frac{\langle (\alpha_s/\pi)
G^2 \rangle}{M^4}
\frac{\pi^2}{6}~\frac{(1-2x^2)}{x^2(1-x)^2(1+2x)}
(1-e^{-s_0/M^2})/E_2(\frac{s_0}{M^2})
\ee

\vspace{2mm}
(The factor $(1-e^{-s_0/M^2})$  appears since  gluonic condensate also
contributes to continuum and this contribution should be subtracted).

Consider now the contributions of d=6 operators.

We start from studying the diagrams of Fig.4 and Fig.5, the contribution of
which is proportional to the vacuum mean values of the type $\langle
\bar{\psi}_{\alpha} \psi_{\beta}D_{\rho} G^n_{\mu \nu} \rangle$, ~$\langle
\bar{\psi}_{\alpha} \nabla_D \psi_{\beta} G^n_{\mu \nu} \rangle$,
~$\langle \bar{\psi} \nabla \nabla \nabla \psi \rangle$ ets. All of them are
expressed through $\langle \bar{\psi} \psi \rangle^2$ (see [12]). Here it
should be necessary to make the following remark: since the approach is
inapplicable at $x \approx 1$, then the diagrams of the type of
Fig.2, the imaginary part of which is proportional to $\delta(1-x)$, should
not to be taken into account. But the diagrams of Fig.4 arising by
evolution of the diagram Fig.2, should not be taken into account too. The
diagrams which contribute to intermediate region $x$ (and not being
evolution of the corresponding non-loop ones) are given in Fig.5 (cf.
Fig.5b in [12]). These diagrams are one-loop ones and for this reason it may
be expected that their contribution will be dominant among contributions of
d=6 operators. The contribution of Fig.5 diagram into sum rules for
quark distribution in proton appeared to be equal to:

\newpage

\be
\frac{u(x)_{\alpha_s \langle \bar{\psi} \psi \rangle^2}}{u_0(x)} =
\frac{\alpha_s a^2 (215-867x+172x^2+288(1-x) ln 2)}{M^6 \cdot 81\pi
\cdot 8x(1-x)^3(1+8x)} \frac{1}{E_2(s_0/M^2)}
\ee

\vspace{3mm}

\be
\frac{d(x)_{\alpha_s \langle \bar{\psi} \psi \rangle^2}}{d_0(x)} = -
\frac{\alpha_s a^2(19-43x+36x^2)}{M^6 81 \pi x(1-x)^3
(1+2x)}~\frac{1}{E_2(s_0/M^2)},
\ee
where

\be
a = -(2 \pi)^2 \langle 0 \vert \bar{\psi} \psi \vert 0 \rangle
\ee

Besides the considered above contribution of the quark condensate square
$\alpha_s \langle 0 \vert \bar{\psi} \psi \vert 0 \rangle^2$, there exist
contributions of d=6 operators described by two-loop diagrams.
Examples of these diagrams are shown in Fig.6. The contributions of these
diagrams are expressed via $\alpha^2_s \langle 0 \vert \bar{\psi} \psi \vert
0 \rangle^2$ and $\langle 0 \vert g^3 f^{abc} G^a_{\mu \nu} G^b_{\nu
\lambda} G^c_{\lambda \mu} \vert 0 \rangle$. The diagrams of the first
type are suppressed in comparison with (27), (28) by $\alpha_s/\pi$ and by
numerical factor $\sim 1/2 \pi$, because they are two loop diagrams. The
diagrams of the second type, proportional to
and $\langle 0 \vert g^3 f^{abc} G^a_{\mu
\nu} G^b_{\nu \lambda} G^c_{\lambda \mu} \vert 0 \rangle$ are also
suppressed by two-loop factor $1/2 \pi$, but vacuum expectation value of
$G^3$ operator is poorly known: there is only instanton estimate [21]

\be
\langle g^3 f^{abc} G^u G^b G^c \rangle = -\frac{48
\pi^2}{5}~(1/\rho^2_c) \langle 0 \vert (\alpha_s/\pi) G^2 \vert 0
\rangle \ee where $\rho_c$ is the effective radius of instanton.
The magnitude of $\rho_c$ is not well known:  various estimations
result in essenially different values, varying from 1/3 [22] to 1
fm [21], Also, one may have doubts [23] if instantons
quantitatively describe physical processes at the scale $\sim 1
GeV^2$, where we are working. Moreover,  some of these diagrams
are infrared divergent (for examples see Fig.6b).

For these reasons we restrict ourselves to the estimate of $\langle G^3
\rangle$ contribution only. Such estimate shows, that it may be essential in
the domain of small $x \la 0.2$ and is positive in both cases -- for $u$-
and $d$-quarks.

The final result for valence quark distribution in proton are of the form

$$
xu(x) = \frac{M^6 e^{m^2/M^2}}{\bar{\lambda}^2_N} 2x
(1-x)^2(1+8x)E_2(\frac{s_0}{M^2}) \left \{ \Biggl [1 +
\frac{u^{LO}(x,Q^2_0)}{u_0(x)} \Biggr ] + \right.
$$
\be
\left. + \frac{1}{u_0(x)} \Biggl [ u(x)_{\langle G^2 \rangle} +
u(x)_{\alpha_s \langle \bar{\psi} \psi \rangle^2}
\Biggr ]
\right \}
\ee

\vspace{3mm}


$$
xd(x) = \frac{M^6 e^{m^2/M^2}}{\bar{\lambda}^2_N} 2x
(1-x)^2(1+2x)E_2(\frac{s_0}{M^2}) \left \{ \Biggl [1 +
\frac{d^{LO}(x,Q^2_0)}{d_0(x)} \Biggr ] + \right.
$$
\be
\left. + \frac{1}{d_0(x)} \Biggl [ d(x)_{\langle G^2 \rangle} +
d(x)_{\alpha_s \langle \bar{\psi} \psi \rangle^2}
\Biggr ]
\right \}
\ee

\newpage

\begin{center}
{\bf 5. Valence $u$- and $d$-quark distributions at intermediate $x$.}
\end{center}

\vspace{3mm}

We are now in a position to calculate numerically valence $u$- and
$d$-quark distributions using eqs.(31),(32). The following values
of the parameters were taken: $\tilde{\lambda}^2_N = 2.1 GeV^6$,
$s_0 = 2.3 GeV^2$ [24], $\lambda_{QCD} = 250 MeV$. The latter is
the effective one-loop value, which gives the same $\alpha_s(Q^2)$
as many-loop calculations at low $Q^2 = 1-5 GeV^2$ and
particularly, $\alpha_s(m^2_{\tau}) = 0.355$ found in [23]. The
parameter $\alpha_s \langle 0 \vert \bar{\psi} \psi \vert o
\rangle^2$ was varied from $8.10^{-5} GeV^6$ to $2.2.10^{-4}
GeV^6$, i.e, $\alpha_s a^2$ varied from $0.13 GeV^6$ to $0.34
GeV^6$. (The lower limit is the old value [25], the upper limit
corresponds to the recent determination of this parameter from
$\tau$-decay data [26],[27]). The important parameter is the value
of gluon condensate $b = \langle 0 \vert (\alpha_s/\pi) G^2_{\mu
\nu} \vert 0 \rangle$. We allow it to vary from $0$ up to $0.012
GeV^4$. (The recent determination of $\langle 0 \vert
(\alpha_s/\pi) G^2_{\mu \nu} \vert 0 \rangle$ from the analysis of
the $\tau$-decay data results in:  $\langle 0 \vert (\alpha_s/\pi)
G^2_{\mu \nu} \vert 0 \rangle = 0.009 \pm 0.007 GeV^4$ [23], the
value $0.012 GeV^4$ is the old result [28]).  All results are
presented at $Q^2_0 = 5 GeV^2$.

Let us start with studying of the Borel mass dependence of quark
distributions at various $x$. (Fig.7; the chosen values of the parameters
are: $a^2 = 0.34 GeV^6$, $b = 0.006 GeV^4$ -- our favourite values -- see
below.) As is seen from Fig.7, for $u$-quark the stability in $M^2$ is good
or satisfactory at $0.1 \la x \la 0.5$, at $x = 0.6$ the $M^2$ dependence is
rather strong. For $d$-quark distribution the stability interval is
narrower; $0.25 \la x \la 0.55$.

Check now how much are the contributions of power corrections in comparison
with $d=0$ OPE term (bare loop). For $u$-quark the applicability domain at
low $x$ is limited by gluon condensate contribution -- it comprises about
$30\%$ at $x=0.15$ (and $M^2=1.1 GeV^2$). At the same $x=0.15$ the
contribution of $\langle G^3 \rangle$ term becomes large, although
uncertain. At large $x$ the limits for $u$-quark come from perturbative
corrections, they comprise about 40\% at $x=0.65$. For
$d$-quark the corresponding limits are $0.25 \la x \la 0.60$. (All the
mentioned above values are given for $\alpha_s a^2=0.34 GeV^6$, $b = 0.006
GeV^4$, $M^2 = 1.1 GeV^2$).

Fig.8a,b shows the dependence of $u(x)$, $d(x)$ on the magnitude
of gluon condensate. One can see from Fig.8, that the data
following from the analysis of deep inelastic scattering are best
described at $b=0.006 GeV^4$. The zero value of gluon condensate
cannot be excluded with certainty, but higher values, $b \ga 0.012
GeV^4$ give much worse description of the data. The curves in
Fig.8a,b were calculated at $\alpha_sa^2=0.34 GeV^6$ (and $M^2 =
1.1.GeV^2$).The dependence of quark distributions on the magnitude
of quark condensate square times $\alpha_s$ --  $\alpha_s a^2 =
(2\pi)^4 \alpha_s \langle \vert \bar{\psi} \psi \vert 0 \rangle^2$
is plotted in Fig.9a,b (at $b=0.006 GeV^4$). The best fit is
obtained at $\alpha_s a^2 = 0.34 GeV^6$.  Therefore, as follows
from the analysis, our favourite values of gluon condensate and
$\alpha_s$ times quark condensate square are

$$
\langle 0 \vert \frac{\alpha_s}{\pi} G^2_{\mu \nu} \vert 0 \rangle = 0.006
GeV^4
$$
\be
\alpha_s \langle 0 \vert \bar{\psi} \psi \vert 0 \rangle^2 = 2.2.\cdot
10^{-4} GeV^6
\ee
which agree with the recent result from the $\tau$-decay analysis [23],
[26], [27].

Besides, the uncertainties arising from the spread of possible values of $b$
and $a^2$, there are additional sourses of errors -- the uncertainty in
$\bar{\lambda}^2_N$ and those coming from continuum contribution. The value
of $\bar{\lambda}^2_N = 2.1 GeV^6$ was found in [24] by the best fit of the
proton QCD sum rules. However, in [24] the old value of $\alpha_s \langle 0
\vert \psi \psi \vert 0 \rangle^2 = 8.10^{-5} GeV^6$ was used and $\alpha_s$
corrections were not accounted. This may result in $10-20\%$ deviation of
$\bar{\lambda}^2_N$ from the above accepted value.  Continuum contribution
comprises about $30\%$ of the total in $u$-quark case and about $60\%$ in
$d$-quark case.  So, one may estimate possible errors from uncertainty of
continuum model as 10\% and 20\%, correspondingly, for $u$- and $d$- quarks.
Therefore, our estimates of total errors are: for $u$-quarks distributions,
in the middle of $x$-interval -- $0.25< x <0.45$, about 25\%, at the ends of
intervals -- $x=0.15$ and $0.65$ -- about 50\%; for $d$ quarks
distributions, in the middle of $x$-interval -- $0.3 < x < 0.45$ about 30\%,
at the ends of intervals -- $x = 0.25$ and $0.55$ by a factor of 2. Taking
into account these errors, the agreement with $u$- and $d$-quark
distributions, found in [4] from experimental data (solid curve in Fig.8) is
satisfactory. It should be mentioned, that the account of $\langle G^3
\rangle$ terms would improve the agreement in the domain of low $x$.

A few additional remarks are in order. In Figs.8,9   we performed
the comparison with valence quark distributions found from hard
processes, we used only one group results [4]. We limited our
comparison to these data not because we consider these results as
more confident or reliable, than those of other groups [5-9].
Quite opposite, we believe, that the precision of new analysis
(see especially [29],[30]) is better than the older ones. The
reason is that theoretical errors in determination of $u$- and
$d$- quark distribution  exceed the differences in various
treatment of the data -- all of them are in the limit of our
theoretical uncertainties. We chosed for comparison the LO results
of [4], because our perturbative calculations are done also in LO.
Our results can be compared directly with experimental data by
considering the difference

\be
F^p_2(x)-F^n_2(x) = \frac{1}{3} [u(x) - d(x)]_{val} + \frac{1}{3}
[\bar{u}(x) - \bar{d}(x)] \ee $F^p_2(x)- F^n_2(x)$ was measured by
NMC [31]. At $x > 0.30$ one may expect, that sea quark
contribution -- the last term in (36) -- is negligible and the
data can be compared with valence quark distributions. The result
is that at $0.3 < x < 0.55$ the theoretical curve (at $b=0.006$
GeV$^6$) is about 30-40\% higher than the data points.  Such
disagreement is in the limit of estimated errors. (At $b=0$ the
agreement is better.)

\vspace{1cm}

{\bf 6. Conclusion}

\vspace{3mm}

The distributions of valence $u$- and $d$-quarks at low $Q^2$ and
intermediate $x$ were theoretically calculated basing on the first
principles of QCD: perturbation theory and operator product expansion (OPE).
No experimental data and no fitting parameters were used. New technique,
suggested in ref.15, was exploited: the double Borel transformation in
virtualities of two hadronic currents in 4-point function eq.(1), which
allows one to kill the background non-diagonal transition amplitudes. In OPE
the operators of dimension 4 (gluon condensate) and dimension 6 (quark
condensate square times $\alpha_s$) were accounted.
The theoretical analysis of the obtained valence
quark distributions at $Q^2 = 5 GeV^2$ showed, that $u$-quark distribution
is reliable at $0.15 < x  < 0.65$, its accuracy is about 25\% in the middle
of this interval and decreases to 50\% at the ends of interval; $d$-quark
distribution is reliable at $0.25 < x < 0.55$ with an accuracy of about 30\%
in the middle and is given by a factor of 2 at the ends of interval. In the
limit of these accuracies the theoretically calculated valence quark
distributions are in agreement with those found from deep inelastic
scattering and other hard processes data. The account of $G^3$ contribution,
probably, improves the agreement, especially at small $x$.

\newpage

{\bf Acknowledgement}

\vspace{3mm}

The authors are thankful to Wu-Ki Tung for his kind presentation of the CTEQ
analysis data. The research described in this publication was made possible
in part by Award No.RP2-2247 of the US Civilian and Development Foundation
for the Independent States of Former Soviet Union (CRDF), by INTAS Grant
2000, Project 587 and by the Russian Found of Basic Research, Grant No.
00-02-17808.

\newpage

\begin{figure}
\epsfxsize=10cm
\epsfbox{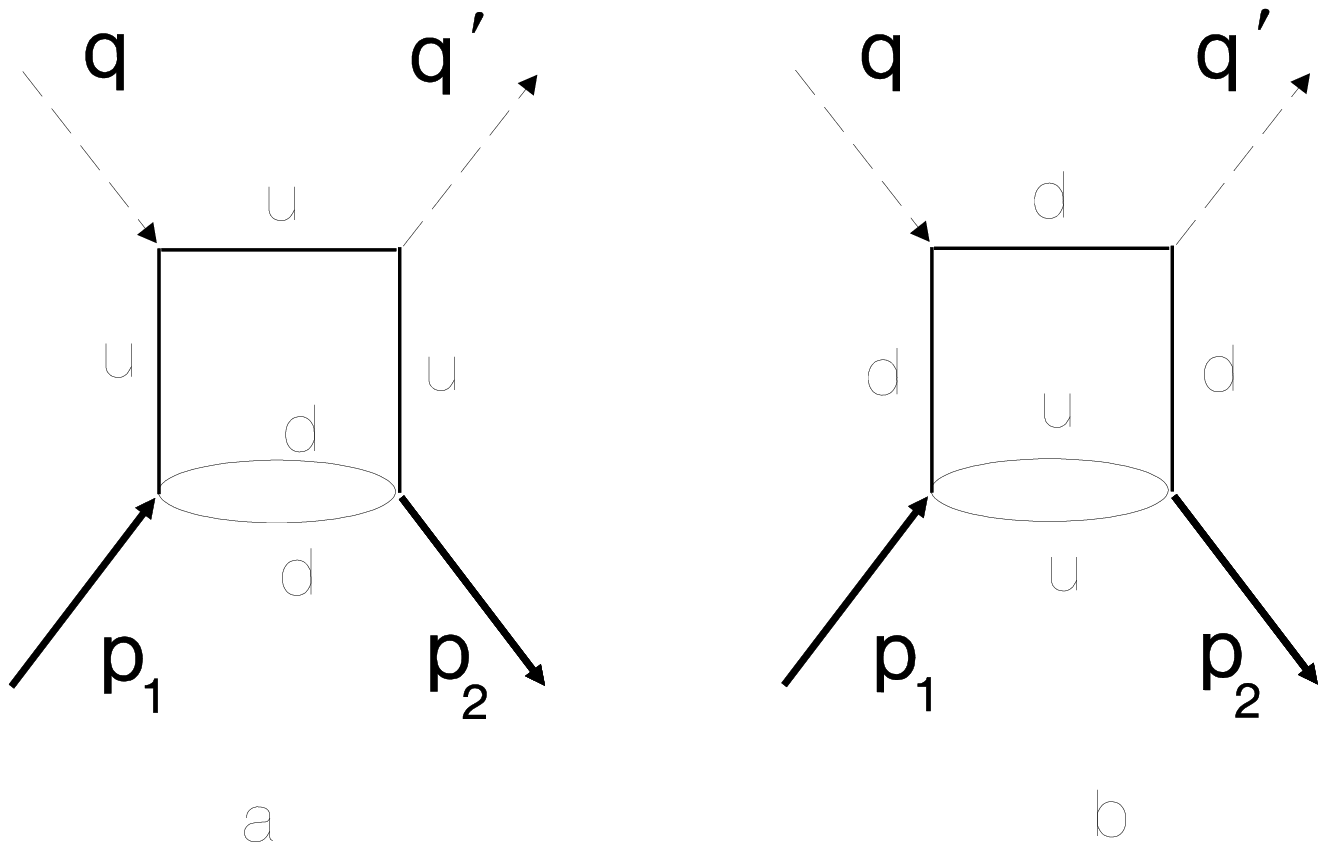}
\caption{Bare loop  diagrams, corresponding to unit  operator
contribution for $u$- and $d$-quarks (respectively, a) and b)).}
\end{figure}

\begin{figure}
\epsfxsize=10cm
\epsfbox{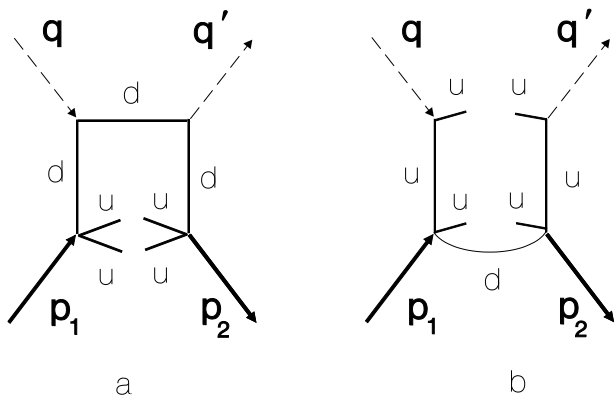}
\caption{Examples of non-loop diagrams $(d=6)$ for $d$-
and $u$-quarks (respectively a),b)).}
\end{figure}

\begin{figure}
\epsfxsize=10cm
\epsfbox{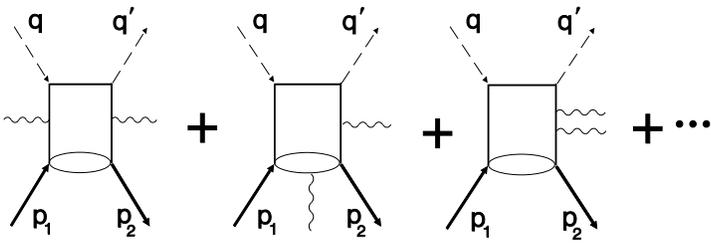}
\caption{ Examples
of diagrams for $d=4$ operator contribution,  wavy lines
correspond to gluon.}
\end{figure}

\begin{figure}
\epsfxsize=10cm
\epsfbox{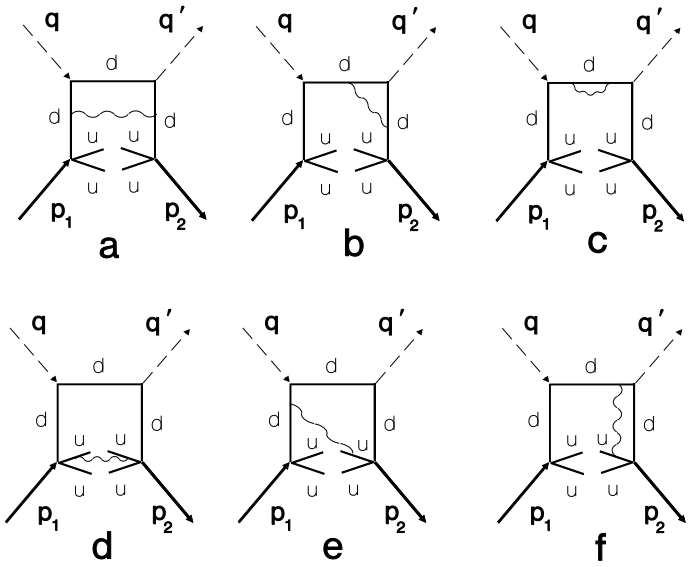}
\caption{Diagrams for $d=6$
contribution, which are  treated as perturbative correction to
non-loop ones and should be omitted. Other notations as in Fig.3.
(For diagrams b),~d),~e),~f) symmetrical diagrams are not shown).}
\end{figure}

\begin{figure}
\epsfxsize=10cm
\epsfbox{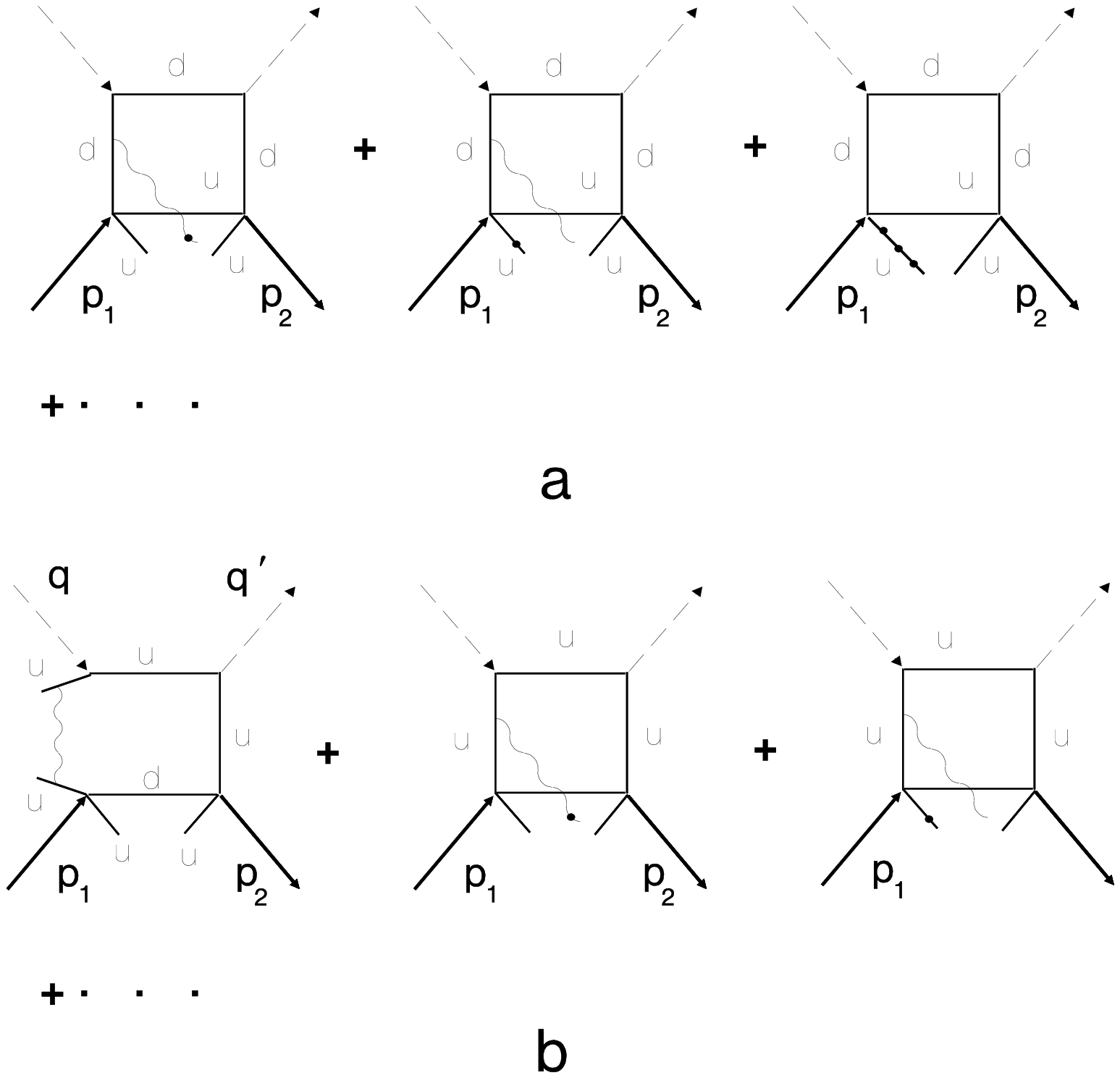}
\caption{Examples of $d=6$ diagrams, which are taken into
consideration, non-labelled quark propagator means that it can be
both of $u,d$-quarks,  other notations -- as in Fig.3.}
\end{figure}

\begin{figure}
\epsfxsize=10cm
\epsfbox{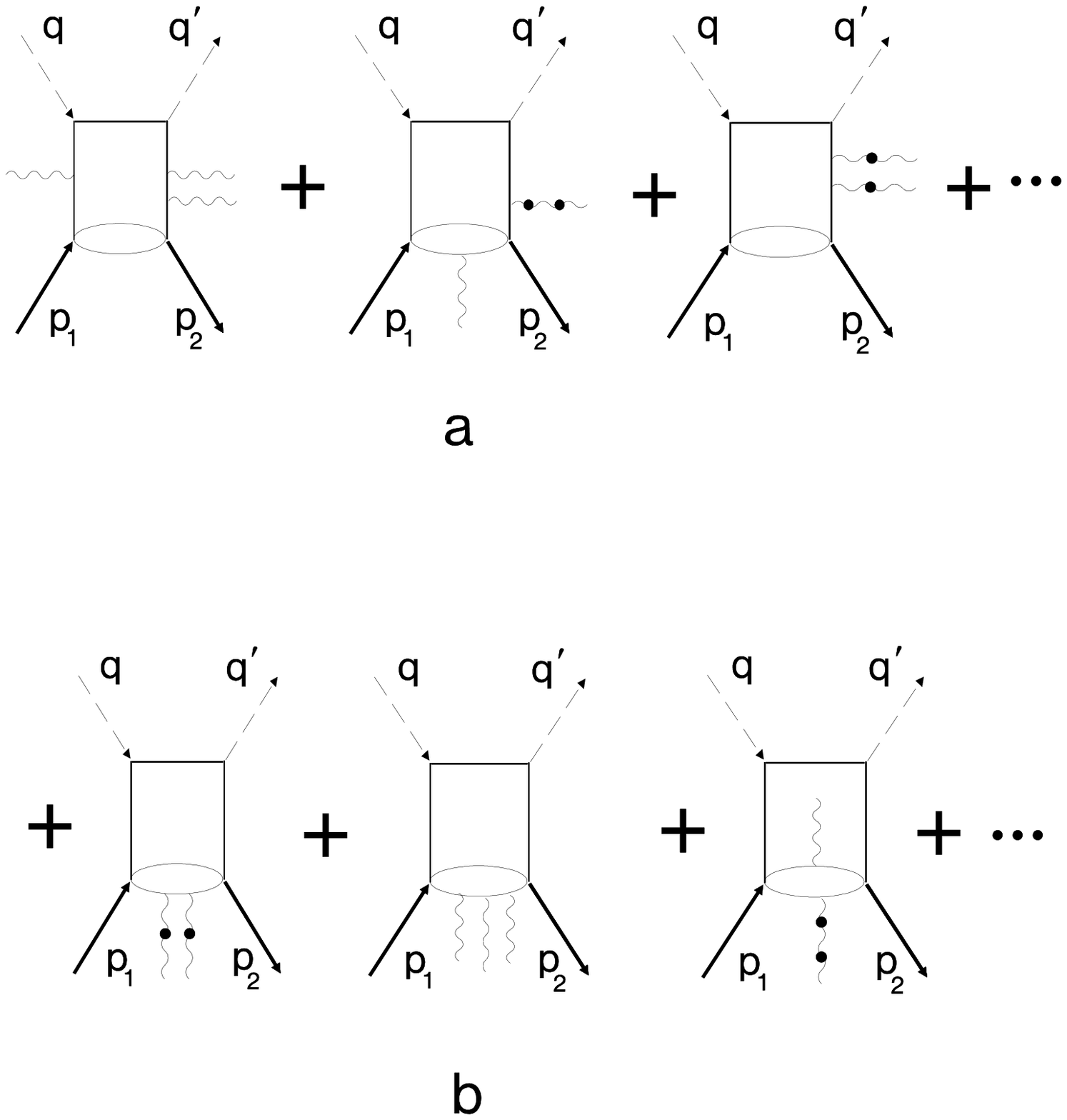}
\caption{Examples of two-loop diagrams of $(d=6)$,(a,b). Diagrams
b) are examples of infrared divergent diagrams.}
\end{figure}

\begin{figure}
\epsfxsize=10cm
\epsfbox{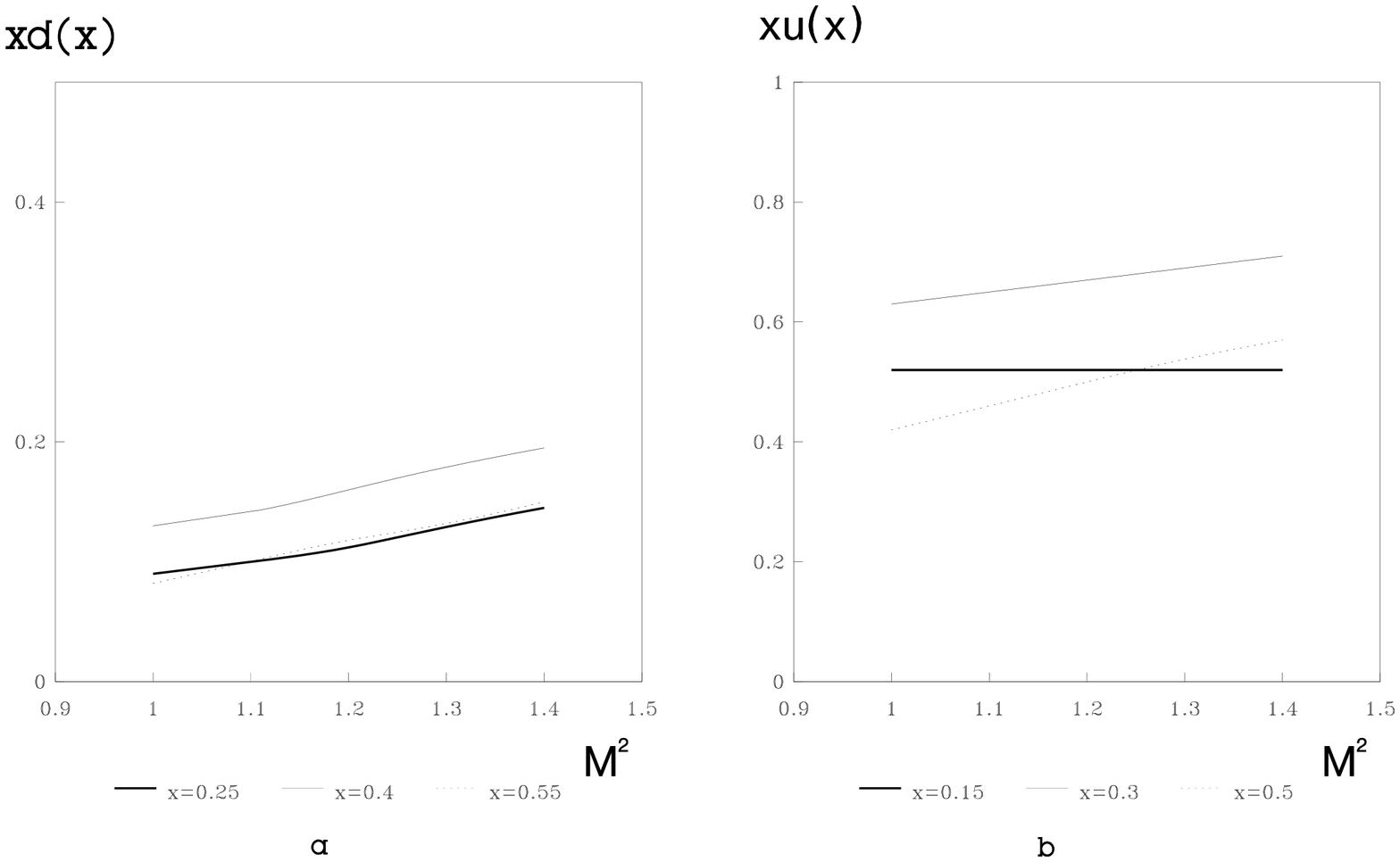}
\caption{ Borel mass dependence of sum rules for $u$- and
$d$-quark distributions at various  $x$.}
\end{figure}

\begin{figure}
\epsfxsize=10cm
\epsfbox{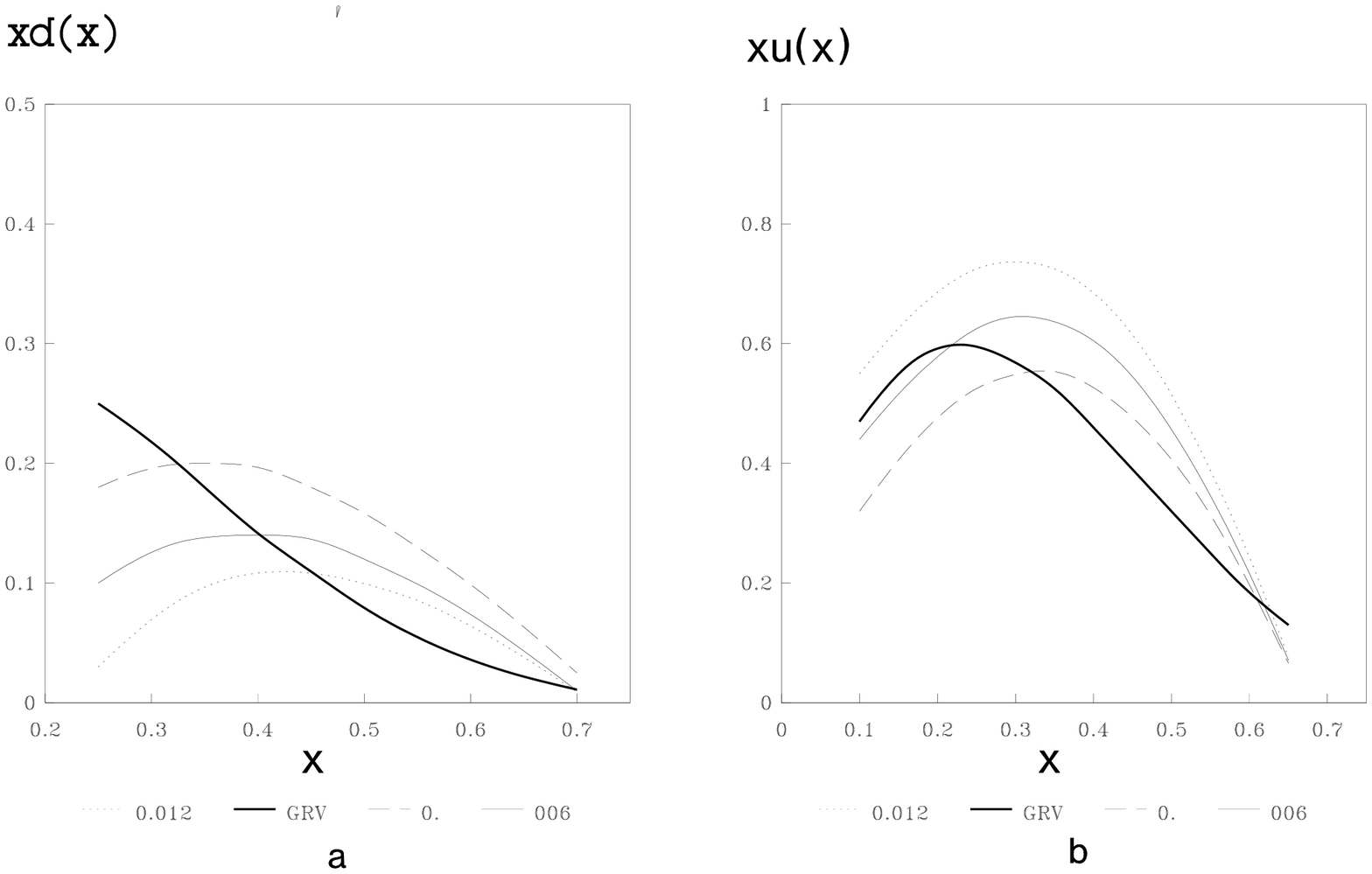}
\caption{ $d$- and $u$-quark distribution at various values of gluon 
condensate ($b=0.012, 0.06$ and 0 GeV$^4$, respectively  dotted, solid 
and dashed lines). Thick solid line corresponds to the results of [4].}
\end{figure}

\begin{figure}
\epsfxsize=10cm
\epsfbox{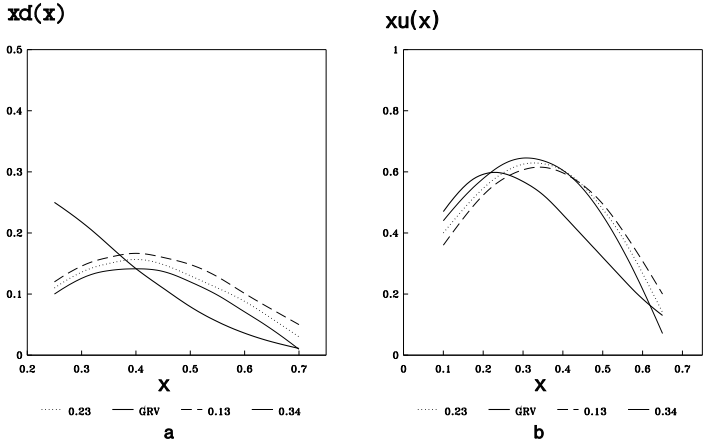}
\caption{ $d$- and $u$-quark distribution at
various values of quark condensate $\alpha_s a^2=0.13,~0.23,~0.34$
GeV$^6$ in comparison with the [4] result (the curve denoted as
GRV).}
\end{figure}

\end{document}